%% file: 00_main.tex
\documentclass[conference,letterpaper]{IEEEtran}
\IEEEoverridecommandlockouts
\usepackage{cite}
\usepackage{amsmath,amssymb,amsfonts}
\usepackage{algorithmic}
\usepackage{graphicx}
\usepackage{textcomp}
\usepackage{listings}
\usepackage{xcolor}
\def\BibTeX{{\rm B\kern-.05em{\sc i\kern-.025em b}\kern-.08em
    T\kern-.1667em\lower.7ex\hbox{E}\kern-.125emX}}
\DeclareUnicodeCharacter{02BB}{{okina}}
\begin{document}

\title{A Decentralized Authorization and Security Framework for Distributed Research Workflows
*\\

\thanks{* This material is based upon work supported by the National Science Foundation Office of Advanced CyberInfrastructure, Collaborative Proposal: Frameworks: Project Tapis: Next Generation Software for Distributed Research (Award \#1931439)}
}

\makeatletter
\newcommand{\linebreakand}{%
  \end{@IEEEauthorhalign}
  \hfill\mbox{}\par
  \mbox{}\hfill\begin{@IEEEauthorhalign}
}
\makeatother

\author{\IEEEauthorblockN{Richard Cardone}
\IEEEauthorblockA{\textit{Texas Advanced Computing Center} \\
\textit{University of Texas at Austin}\\
Austin, TX, USA \\
rcardone@tacc.utexas.edu}
\and
\IEEEauthorblockN{Smruti Padhy}
\IEEEauthorblockA{\textit{Texas Advanced Computing Center} \\
\textit{University of Texas at Austin}\\
Austin, TX, USA \\
spadhy@tacc.utexas.edu}
\and
\IEEEauthorblockN{Steven Black}
\IEEEauthorblockA{\textit{Texas Advanced Computing Center} \\
\textit{University of Texas at Austin}\\
Austin, TX, USA \\
scblack@tacc.utexas.edu}
\linebreakand %
\IEEEauthorblockN{Sean Cleveland}
\IEEEauthorblockA{\textit{Information Technology Services} \\
\textit{University of Hawai‘i - System}\\
Honolulu, HI, USA \\
seanbc@hawaii.edu}
\and
\IEEEauthorblockN{Joe Stubbs}
\IEEEauthorblockA{\textit{Texas Advanced Computing Center} \\
\textit{University of Texas at Austin}\\
Austin, TX, USA \\
jstubbs@tacc.utexas.edu}
}


\maketitle

\begin{abstract}
Research challenges such as climate change and the search for habitable planets increasingly use academic and commercial computing resources distributed across different institutions and physical sites. Furthermore, such analyses often require a level of automation that precludes direct human interaction, and securing these workflows involves adherence to security policies across institutions. In this paper, we present a decentralized authorization and security framework that enables researchers to utilize resources across different sites while allowing service providers to maintain autonomy over their secrets and authorization policies. We describe this framework as part of the Tapis platform, a web-based, hosted API used by researchers from multiple institutions, and we measure the performance of various authorization and security queries, including cross-site queries. We conclude with two use case studies – a project at the University of Hawaii to study climate change and the NASA NEID telescope project that searches the galaxy for exoplanets.
\end{abstract}


\input{01_introduction.tex}
\input{02_background.tex}
\input{03_design.tex}

\input{04_performance.tex}
\input{05_usecase.tex}
\input{06_related_work.tex}
\input{07_conclusion_futurework.tex}
\bibliographystyle{IEEEtran}
\bibliography{IEEEabrv,tapis_security}

\end{document}

%% file: 01_introduction.tex
\section{Introduction}
\label{section:intro}
Today’s computationally intensive research workloads often involve workflows that span the cyberinfrastructure (CI) of multiple commercial, academic and governmental facilities.  These facilities represent distinct physical, geographic and administrative domains.  While processing across sites and institutions brings more resources to bear on a problem, it also increases the complexity of workload management and security.  

For example, climatology studies often require multi-step analyses. The first step aggregates data collected in real-time by sensors on weather stations. This step typically involves quality control and anomaly detection over the large quantity of data being collected, making it advantageous to perform this step on resources co-located with the data. If the analysis detects outliers, a second step is triggered to execute mathematical simulations capable of high-precision forecasting. Some simulations require high-performance computing resources far from weather stations, e.g., cloud simulations based on Lagrangian Particle Tracking (\cite{UHclimate}).

In this paper, we concentrate on the security implications of running cross-site, cross-institutional workloads.  We present the decentralized authentication, authorization and secrets management subsystem of Tapis \cite{tapisRefJstubbs2021}, a web-based microservice platform for executing workloads on high performance computers (HPC), cloud and private cluster systems.  Clients interact with Tapis using REST APIs, which allows the use of standard HTTP proxies and load balancers for request routing. 

A single Tapis instance comprises a primary site and zero or more associate sites.  Each site defines one or more tenants, and each tenant has its own users and identity provider.  A distributed role-based access control (RBAC) and permissions service provides the most basic type of authorization, while the notion of shared contexts allows for more powerful collaboration patterns.  Tapis also manages passwords, keys and other secrets using Hashicorp’s Vault \cite{vault} as a backend.  All aspects of security can be managed at the site level, giving individual institutions the ability to enforce their own security policies.  Tapis implements a unique request routing algorithm to support secure intersite interactions.

We show how Tapis’s multi-site, multi-tenancy, microservice architecture achieves a high level of flexibility and extensibility.  Authentication can be extended by incorporating new identity providers in new tenants.  Within a tenant, users can be dynamically assigned and unassigned roles, permissions, secrets and shared contexts.  New microservices that weren’t anticipated when the system was designed can be added anytime.  

Associate sites have the additional flexibility of choosing which Tapis services they run locally and which they rely on the primary site to provide.  The goal is to minimize the resource and administrative overhead at associate sites, making associate site set up and operation as lightweight as possible.  The two reasons for establishing an associate site rather than simply interacting with the primary site are to (1) improve performance and (2) maintain local control of security policies and secrets.  Performance can sometimes be improved by moving the execution to the data.  In addition, some institutions require that authentication and authorization occur behind their firewall or that secrets never leave their site.  


\textbf{Contributions}
This paper makes these contributions:
\begin{itemize}
\item Describes the security aspects of Tapis’s multi-site \mbox{microservice} architecture.
\item Implements an extension to the Apache Shiro \cite{shiro} \mbox{permission} model.
\item Implements a request routing algorithm that enforces site-level security.
\item Introduces a way to share security contexts across \mbox{microservices}. 
\end{itemize}

%% file: 02_background.tex
\section{Background}

\subsection{General Tapis Overview}
Tapis is a cloud-hosted, HTTP API framework for automating interactions with storage and computing resources. Investigators use Tapis to manage data and metadata as well as execute programs on high-performance, high-throughput and cloud computing resources. Tapis records the activities taken in the platform and makes this history available through various endpoints, allowing analyses to be repeated more easily and thus aiding in the reproducibility of results. Additionally, users leverage Tapis's fine-grained permissions model to facilitate collaboration: objects registered with Tapis start out as private to a single user (the ``owner") but they can be shared with other individuals or with entire research communities.


Tapis originated in the iPlant Collaborative project as the Foundation API in 2008, and three major open-source versions have been released. To date, over 50,000 unique users across dozens of projects funded by agencies including CDC, DARPA, NASA, NIH and NSF have leveraged Tapis. This paper focuses exclusively on Tapis v3, the most recent version of the platform, funded by NSF in 2019. Tapis v3 made its 1.0 production release in July of 2021.

Tapis leverages a \textit{microservice} architecture where each service is responsible for one or more primary Tapis abstractions, and services coordinate with each other to accomplish the end-user's objective. Today, a total of 16 independent services comprise Tapis v3, up from 12 at its initial production launch in 2021.

\subsection{Data Management and Code Execution – Systems, Apps, Files and Jobs}

The Systems, Apps, Files and Jobs (SAFJ) APIs provide data management and code execution functionality on external resources. Users define \textit{systems}, the foundational abstraction, by providing the information Tapis needs to connect to and communicate with the storage and/or computing resources.  Tapis systems include HPC clusters, virtual machines running in academic or commercial clouds, campus and/or lab servers, or a number of other supported resource types, such as AWS S3 \cite{s3} buckets or iRODS \cite{irods} servers. Similarly, users define \textit{apps} which contain the research analysis software they wish Tapis to execute. Once systems and apps have been defined, the Files API allows for synchronous data ingest and download as well as asynchronous data transfer between systems, while the Jobs API is used to invoke instances of applications.

\subsection{Real-time and Event Programming – Streams, Functions and Notifications}
While the SAFJ APIs enable data management and code execution on external resources, Tapis provides a set of services that facilitate data analysis on its own internal resources. These services provide higher throughput and lower latency than the SAFJ APIs with the goal of enabling real-time data processing. The Streams API allows projects to ingest measurement data from IoT/sensors into a Tapis-managed time-series database, while the Tapis Actors service enables users to register small functions that execute on Tapis's cloud infrastructure in response to messages. Furthermore, Streams can be configured with \textit{channels} that define thresholds to trigger actions, such as the execution of an actor/function. Finally, the Notifications API allows users to subscribe to other kinds of Tapis events, such as the completion of a job. Together, these services provide the building blocks
to support real-time, event-driven architectures, including automated data analysis pipelines that continually process new data as it arrives, with no human in-the-loop.

\subsection{Experiment Metadata – DBs-as-a-service}
In addition to the Streams API supporting time-series data, the Tapis Meta API provides a managed document store based on MongoDB, while the Tapis PgREST service provides a hosted, multi-tenant API for relational data based on PostgreSQL. Both of these services provide a means for projects to store various kinds of data without managing any servers. For example, the VDJServer project \cite{vdjPaper} uses the Meta API to store and manage billions of records related to immune repertoire analysis.

%% file: 03_design.tex
\section{Sites, Tenancy and Authentication} 
Typically, a Tapis installation involves software components deployed across multiple institutions, referred to as Tapis \textit{sites}, with one site identified as the \textit{primary site} and the others called \textit{associate sites}. Components coordinate across sites in a ``hub-and-spoke" model, where requests can route between an associate sites and the primary site but not from one associate site to another associate site. This multi-site approach allows institutions to run only a subset of the Tapis components on-premise, reducing the overall administrative burden, while at the same time having access to all Tapis functionality for their projects. Additionally, deploying Tapis components within an institution's datacenter may improve their security posture and may also result in performance gains in certain cases; we explore both aspects in subsequent sections.

In Tapis, the concept of a \textit{tenant} creates a logical partition of all platform objects, including authentication (users, tokens, API keys, etc.), authorization (permissions, roles, administrative accounts, etc.), and the various Tapis resources (systems, apps, streams, jobs, functions, etc). Thus, research communities or individual projects assigned their own tenants control which users can authenticate, how objects (datasets, applications, etc.) are shared within the tenant, and other administrative aspects of the platform.  In general, \textit{users in one tenant have no visibility into the resources or actions of users in any other tenant}. 

Each tenant is owned by a single site -- either the primary site or an associate site -- and that site maintains administrative control over security aspects of the tenant, including the RSA keys used to sign tokens. The Tapis Tenants API maintains the list of all tenants in a Tapis installation including various metadata about the tenant, such as URLs associated with the tenant and the current public key used to validate tokens.    

\begin{figure}%
    \centering
    {{\includegraphics[width=10cm]
    {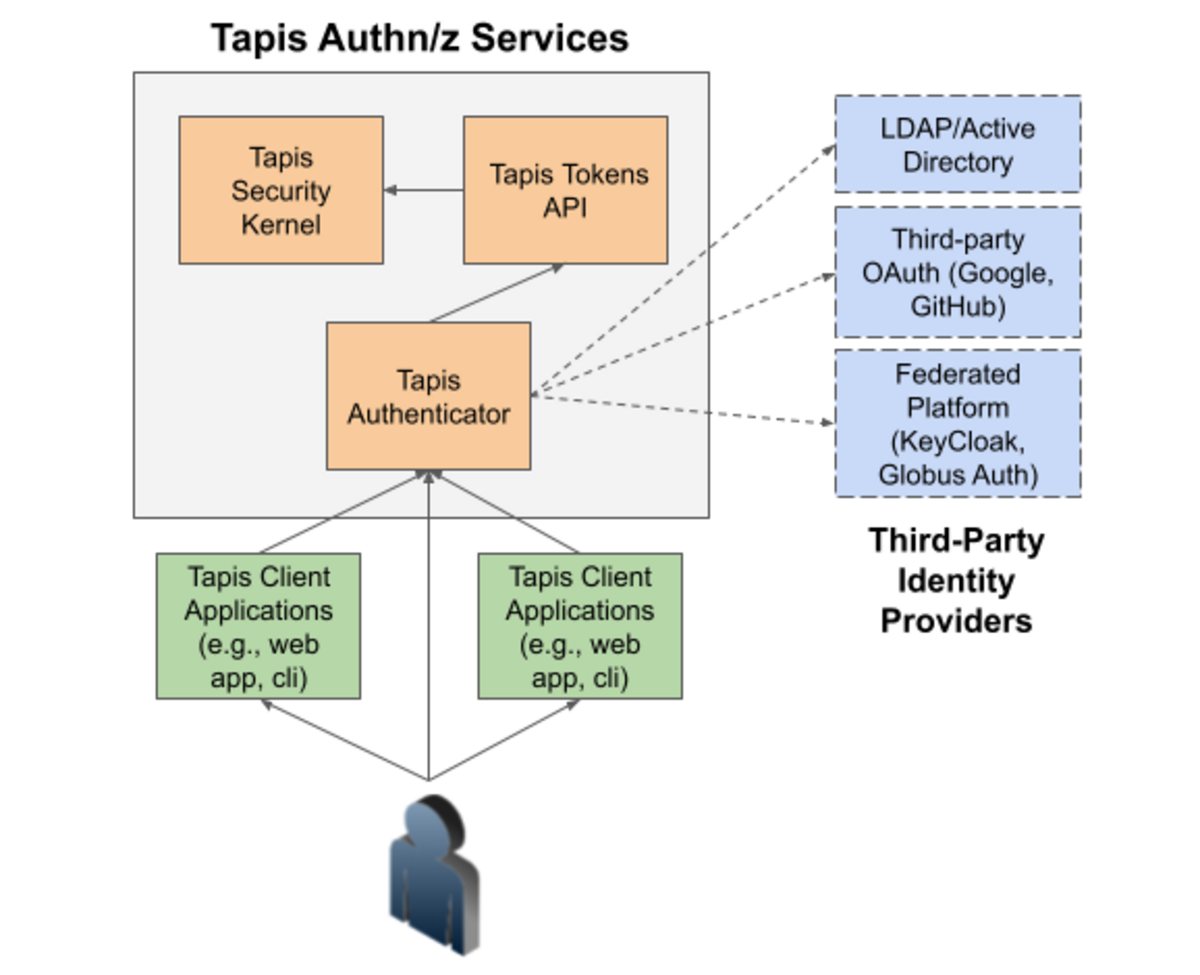}}}%
    \caption{Authentication in Tapis}%
    \label{fig:authn}%
\end{figure}

Tapis divides authentication across two microservices to achieve modularity and flexibility. At the lowest level, the Tokens API issues cryptographically signed JSON Web Tokens (JWT) \cite{jwt} for all subjects in a tenant. Each JWT includes information about the subject represented by the token in the form of claims, including standard claims such as \texttt{sub} which take values of the form \textit{username@tenant}. Tapis assigns an instance of the Tokens API to each tenant in the form of a URL (e.g., https://a2cps.tapis.io/v3/tokens), allowing different tenants to use different instances (or even different implementations) of the service. Additionally, each tenant is configured with a unique public/private RSA key-pair used for signing tokens. Only the Tokens service assigned to the tenant can access the tenant's private key. The other Tapis services use the public key to verify the signatures on JWTs passed in requests made in the tenant. Furthermore, a special role called the \textit{token\_generator} role controls the subjects authorized to create user tokens via the Tokens API for a given tenant. This approach allows tenants to use multiple authenticating services, or \textit{authenticators}, should that be desirable. 

Authenticators authenticate end-users and issue calls to the Tokens API to generate their tokens. Tapis includes a default Authenticator service based on OAuth2 protocols. The Tapis Authenticator can be configured to use LDAP or Active Directory for checking user credentials, or to use a third-party OAuth server, such as Google or GitHub. In the latter case, Tapis client applications (e.g., web applications, mobile applications, CLIs, etc.) use a traditional OAuth grant type (e.g., \textit{authorization code} or \textit{implicit}) directly with Tapis, and the Authenticator in turn manages a separate OAuth flow with the third-party system. In this way, the Tapis Authenticator insulates its client applications from various authentication implementations, providing a single API to any supported authentication mechanism. Tapis Authenticator also supports authentication with identity federation platforms such as Globus Auth \cite{Globus-Auth}, allowing for the use of identity federations such as InCommon \cite{incommon} in a single Tapis tenant. 

Still, projects are free to replace the Tapis Authenticator and/or Tokens APIs with their own, custom services. The other Tapis services only require that requests include a JWT formatted with the expected claims and signed using the private key associated with the tenant. In this way, the Tapis authentication stack allows for extensive customization without modification to any of the other services.

\section{Tapis Security Kernel and Authorization}
\label{sec:tapis-sk}

As shown in Figure \ref{fig:authn}, the Security Kernel (SK), along with Tokens and Authenticator, comprise the three components that make up the Tapis authentication and authorization infrastructure required at each site.  The SK and Tokens instances are only accessible to services running at that site or to users in tenants owned by that site.  Exactly one SK instance runs per site.  SK's two responsibilities are secrets management and authorization support. 

\subsubsection{Secrets based on Vault}
SK uses Hashicorp Vault \cite{vault} to manage secrets for both users and services.  Secrets are segregated by tenant and fall into five categories: service passwords, DB credentials, system credentials, signing keys and user secrets.  A service passes its service password to Tokens to acquire its service JWT.  Tokens calls SK to validate the password before establishing the service's identity (only Tokens can validate service passwords).  DB credentials are used by services to access their persistent stores.  System credentials are used by the Systems service to store user login credentials in SK; Systems only shares these credentials with Files and Jobs.  User secrets are free-form secret data that end-users and services can create.  As previously mentioned, Tokens signs every JWT with a tenant-specific signing key, and services validate requester identity by checking signatures. 

Each site has an \textit{administrative tenant} accessible only to services.  This tenant provides a namespace for secrets and other resources inaccessible to end-users.  In section \ref{section:tapis_bootstrap} (Bootstrapping) we discuss how secrets are initially established; similar processes are used to update secrets.  Utilities to backup and export Vault data run without interrupting Tapis.      

\subsubsection{Basic Authorization}
All SK authorization information is stored in PostgreSQL.  SK provides REST APIs to create, query, update and delete roles, and to assign roles to users.  Minimally, roles have a name, owner and tenant.  The \textit{hasRole} endpoints return whether a user is assigned a particular role.  

Each service manages its own roles in the site administrative tenant.  In addition, each tenant has at least one user assigned the \textit{tenant-admin} role.  Tenant administrators can define roles in their tenant and users can check those assignments.  Every user is assigned their own default role.   

For added flexibility, roles can contain child roles and a role can be the child of any number of parent roles.  These \mbox{parent}/child relationships form a forest of directed acyclic graphs.  When a user is assigned a role, the user is implicitly assigned all children of that role.  Role nesting encourages the creation of fine-grained roles that can be combined into more powerful aggregate roles.        

\subsubsection{Extended Shiro Permissions}
Roles can also contain permission strings as defined in Apache Shiro \cite{shiro}.  Permissions allow owners of resources to grant fine-grained access to their resources.  Services can choose to support transitive permission granting in which a grantor grants permission on resources previously granted to them, though no service currently does.    

Permissions are colon separated strings that conform to an agreed upon schema.  For example, below are three permissions defined by the Systems service that conform to the schema \\"\textit{service:tenant:operation:systemId}".
\begin{lstlisting}
    a) systems:tacc:read:stampede2
    b) systems:cyverse:*:frontera
    c) systems:a2cps:read,modify:corral
\end{lstlisting}
A grantee of permission a) in the \textit{tacc} tenant has read access to the \textit{stampede2} system.  A grantee of b) in the \textit{cyverse} tenant has all access to \textit{frontera}, and a grantee of c) in the \textit{a2cps} tenant has \textit{read} and \textit{modify} access to \textit{corral}.  

SK's \textit{isPermitted} APIs determine if a required permission is implied by any permission assigned to a user.  For example, no above permission implies d), whereas b) implies e).
\begin{lstlisting}
    d) systems:tacc:modify:stampede2
    e) systems:cyverse:exec:frontera
\end{lstlisting}

Tapis extends the standard Shiro permission matching just described by recognizing hierarchical structures such as file paths.  Consider the follow permission string:
\begin{lstlisting}
    files:tacc:read:sys1:/home/bud/data
\end{lstlisting}
A grantee assigned this permission in the \textit{tacc} tenant can read any file in the subdirectory rooted at \textit{/home/bud/data} on \textit{sys1}.   

\section{Advanced Authorization (Sharing)} \label{section:shares} 
Though roles and permissions provide well understood access control to individual Tapis resources, they cannot coordinate access to multiple resources at once.  Consider the directive, "let Alice execute MyApp", which, once granted, should allow Alice to run MyApp on some system, accessing the required input and output files as needed.  Using only roles and permissions, however, one would have to individually grant access to every system and path referenced in the application definition.  These access controls would require maintenance if the definition changed, necessitating careful bookkeeping to avoid invalidating access when resources are referenced by multiple applications.  

To allow the coordinated sharing of resources, and to avoid the complexity and overhead of managing multiple, fine-grained access controls, SK implements the concept of distributed shared contexts.  These contexts only share resources under certain circumstances, the main example being \textit{\textbf{Shared Application Contexts (SAC)}}.  SACs allow resources specified in application definitions to be shared \textit{only during job execution}.  These resources include execution and storage systems along with their input and output file paths.  

When a user submits a job the Jobs service retrieves the application definition, which may indicate that the job should run in a SAC for that user.  Running a job in a SAC means that Jobs, Files and Systems can use the grantor's authorizations for certain resources specified in the application definition.  The shared context is implemented by passing SAC information on Files and Systems calls that access the shared resources.  Even when running in a SAC, however, the underlying hosts' authorizations are still in effect.  For example, hosts still validate the user's login credentials and authorization checks are still applied by Posix, S3 or other storage systems.  Sharing affects Tapis authorization checking but not host authorization.

The true power of sharing is realized when coupled with \textit{templated} system, application and job definitions.  Certain fields in these definitions can contain \textit{template variables} that are dynamically assigned values at runtime.  This allows the runtime context of a shared application to be tailored to the user executing the job.  For instance, shared systems can be accessed as the user executing the job, allowing multiple users to use the same system without sacrificing security or traceability: Tapis authenticates to hosts using each user's credentials, executes programs as that user, and accesses storage as that user.

SK also supports \textit{tenant-public} and \textit{no-authn} sharing.  The former is a way of implicitly sharing resources with all users in a tenant; the latter allows unauthenticated users access to Tapis resources simply by presenting a URL.  No-authn sharing requires specially configured routing and service endpoints.  

\section{Shared Application Context Example}
To make concepts from the last section more concrete, we illustrate running a job in a SAC.  For this discussion, assume that all services are initialized (see section \ref{section:tapis_bootstrap}) and every API call validates the requester's JWT upon receipt. 

Let Alice define an application, \textit{aliceApp}, in tenant \textit{tenant1} that specifies execution system \textit{execSys} and required input file \textit{inputFile} from storage system \textit{storeSys}.  

We define \textit{execSys} with a template variable that dynamically selects the job requester's credentials when accessing the system.  Conversely, we statically define \textit{storeSys} to always use \textit{storeAdmin} credentials when accessing this system. 

Alice shares \textit{aliceApp} with Bob.  To close off possible avenues for privilege escalation, Alice must have the proper access to every system specified in the application definition at share time.  This prevents, for example, Alice from specifying an execution system to which she has no Tapis access but to which her grantees would gain implicit Tapis access.  Share time access checking is necessary if an application specifies an execution system, an archive system or any file input source systems.  These checks typically require SK interactions to query Alice's privileges. 

Once \textit{aliceApp} is successfully shared, Bob submits a request to execute the application without overriding any values defined in the application definition, but by adding archive system \textit{arcSys} and archive directory \textit{arcDir}.

Jobs receives Bob's HTTPS request with Bob's JWT passed in a header.  Jobs validates the JWT's signature using tenant \textit{tenant1}'s public key obtained from the Tenants service.  Jobs requests the \textit{aliceApp} definition from the Apps service, which validates the Jobs service JWT and recognizes that the request is on behalf of Bob in \textit{tenant1}.  Apps discovers that Bob is not the owner of \textit{aliceApp}, so it calls SK to see if the application has been shared with Bob.  SK affirms \textit{aliceApp} has been shared, so Apps returns the application definition to Jobs with the SAC grantor's name (Alice).

Once Jobs recognizes that \textit{aliceApp} is running in a SAC, all calls to Systems and Files include the SAC information so those services can participate in the shared context.  For instance, when Jobs requests the \textit{execSys} definition from Systems, Systems first validates that Alice still has access to \textit{execSys}.  If so, the system definition is resolved for Bob and returned.  Resolution here means that Bob's credentials are passed back to Jobs because of dynamic template resolution.  If Bob had not previously associated his credentials with \textit{execSys}, the job fails.

On the other hand, if Alice no longer has access to \textit{execSys}, then Systems must determine if Bob is authorized by some other means to run jobs on that system.  For example, Bob could be the owner of \textit{execSys}, or could have had it shared by the owner, or could have been granted permissions on it.  If neither Alice nor Bob is authorized on \textit{execSys}, the job aborts.

Checking the SAC grantor's access to SAC resources at runtime and, when necessary, falling back to grantee checking, avoids anomalies that can occur when access privileges change over time.  Specifically, grantors can lose access to a resource if the resource owner revokes access or, if the grantor was the owner, ownership is transferred.  Without runtime checks, a resource owner could not prevent SAC grantees from accessing a resource after a SAC grantor's access was revoked.

At some point Jobs stages \textit{inputFile} from \textit{storeSys} to a directory on \textit{execSys} by issuing a transfer request with SAC information to Files.  Files, in turn, calls Systems with the SAC information to retrieve the \textit{storeSys} definition.  Systems performs the same runtime authorization checks as just described for \textit{execSys}.  Once authorized, Systems always returns \textit{storeAdmin}'s credentials because templating was not used in \textit{storeSys}'s definition.  Since \textit{storeSys} and \textit{execSys} are part of the SAC, Files does not need to check with SK if Bob is authorized to read the input file or write the output file.  It is, however, the file systems on \textit{storeSys} and \textit{execSys} that are the ultimate arbiters of whether the transfers are authorized.  

Because \textit{aliceApp} did not specify an archive system, \textit{arcSys} is not part of the SAC, so standard Tapis authorization checking is required.  Since Bob needs write access to \textit{arcDir} on \textit{arcSys}, SK queries are required unless Bob owns \textit{arcSys}.    

\section{Multi-Site Security} 
An institution's security policies may dictate an on-premise Tapis deployment rather than using an externally hosted Tapis.  An in-house primary site deployment allows complete control over all authentication, authorization and secrets data, all policies that regulate their usage, and the choice and location of services that run.  With this control comes the responsibility of managing a distributed system with multiple services, their databases, message brokers, proxies, etc.

Alternatively, an associate site deployment reduces the local administrative and resource requirements.  An associate site creates its tenants and chooses which services run on-site.  Every associate site runs at least one instance of Tokens and Authenticator and a single SK instance.  The Tapis authentication, authorization and secrets data associated with locally running services resides on-site.    

It's important for some institutions to control when security artifacts are exposed off-site.  For instance, by running Systems locally user credentials to host systems are managed by the resident SK.  If, however, the associate site depends on Jobs or Files running at the primary site, then user credentials will be transmitted off-site during job execution and file transfer.  Running Systems, Jobs and Files locally removes the exposure in this case.  The key point, however, is that some analysis is required to know if an associate site configuration meets an institution's security requirements.  Currently, several primary and associate site Tapis deployments exist at different institutions.  Tapis sites are virtual, so a physical data center can host any number of primary or associate sites.         

\subsection{User Requests, Service Requests, and Request Routing} 

\begin{figure*}%
    \centering
    {{\includegraphics[width=19cm]
    {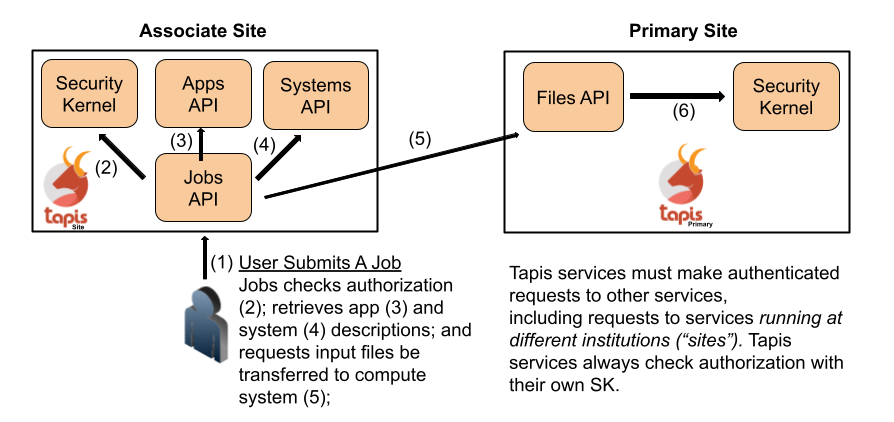}}}%
    \caption{Requests in Tapis}%
    
    \label{fig:requests}%
\end{figure*}

Tapis exposes its HTTPS APIs through various \textit{base URLs}, i.e., the consistent part of the URL within a cloud platform that includes the top-level domain, primary domain, and one or more subdomains. Tapis assigns a unique base URL to each tenant, and these base URLs can be custom domains owned by the project in charge of the tenant, allowing for white labeling, or they can be assigned to a subdomain owned by the Tapis project. For example, by default, a tenant owned by the primary Tapis site running at the Texas Advanced Computing Center will be assigned a base URL of the form \textit{tenant}.tapis.io.

As mentioned previously, every tenant is owned by a single site, either a primary site or an associate site. These sites represent virtual installations of Tapis components and always include the Tapis site router service, an HTTP reverse proxy configured to route requests based on certain rules. For every tenant owned by the site, DNS resolves the tenant's base URL to the site router service installed at the site owning the tenant. When a user makes a request to a Tapis service -- for example, the Apps service -- DNS will resolve the base URL to some site router, and the site router uses the list of services deployed at its site to determine how to route the request: if the request is for a service that runs at that site, the site router will route the request internally to the service; otherwise, the site router will route the request to the primary site.  

TLS is terminated at the last proxy along any route.  These proxies are usually nginx instances running inside a private network, such as a Kubernetes network, which allows non-TLS communication between services inside that private network.        

Organized as a set of independent HTTP microservices, Tapis makes use of \textit{service-to-service} requests (or just \textit{service} requests), i.e., HTTP requests made by one Tapis service to another.  Service requests supply \textit{service JWTs}, whereas user requests supply \textit{user JWTs}. In fact, most user requests trigger multiple service requests. For example, a user request to the Jobs service to invoke an application involves service requests from Jobs to Apps, to retrieve the application definition, to Systems, to retrieve the definition of the execution system, and to Files to transfer inputs for the job to the execution system and archive outputs produced by the job after the execution completes (Figure \ref{fig:requests}). Also, all services make requests to the SK to check authorization.

Service requests make use of base URLs and JWTs in the exact same way that user requests do, and therefore, Tapis requires a method for services to authenticate with each other in some tenant. Tapis uses the concept of an \textit{administrative tenant} to accomplish this goal -- for each Tapis site, an administrative tenant is created so that the Tapis services running at the site can authenticate with the local Tokens API to generate JWTs. Services authenticate with HTTP Basic Authentication \cite{httpBasicAuth} using a password injected into the service's run-time environment configuration at startup. Service password management is part of the Tapis bootstrapping process, described in \ref{section:tapis_bootstrap}. 

Administrative tenants allow services to create and manage Tapis objects in complete isolation from user objects. For example, the Streams service stores project, site, instrument and measurement metadata using the Meta API, and all services make use of authorization data stored in SK. By storing these objects in a separate tenant that cannot be accessed by users, we avoid name collisions.

\subsection{The Tapis Bootstrapping Process}
\label{section:tapis_bootstrap} 
While microservice architectures achieve modularity resulting in extensibility, coordinating the initialization of independent services -- including services at different sites -- presents a challenge. The delicate inter-dependency between the Tenants, Tokens and SK services in particular necessitates a specific deployment order and an administrative utility, \texttt{sk-admin}. Deployment makes use of the following high-level algorithm which we explain in subsequent paragraphs:

\begin{lstlisting}
  1) Deploy Vault;
  2) Run sk-admin -> generate secrets;
  3) If primary_site:
         Deploy Tenants;
     If associate_site:
         Send admin Public Key to Primary;
  4) Deploy SK;
  5) Deploy Tokens;
  6) Deploy Authenticator;
  7) Deploy all other Tapis services;
  
\end{lstlisting}

To initialize a Tapis site -- either a primary site or an associate site -- the Vault service must be deployed before any other service and the \texttt{sk-admin} utility, a command-line Java program, must be executed to generate secrets for the site and store them in Vault. These secrets include: 

\begin{itemize}
    \item[(1)] Public/private RSA key pairs used to sign tokens for all tenants owned by the site.
    \item[(2)] Service passwords for all services running at the site.
    \item[(3)] Database credentials used by services.
    \item[(4)] Additional secrets required by services, such as LDAP bind passwords used by the Authenticator.  
\end{itemize}
Note that \texttt{sk-admin} uses a set of JSON configuration files to determine which secrets it must ensure exist, and it typically runs in an \textit{idempotent} way, meaning, it will create any secrets that do not already exist while ignoring secrets that it created in prior runs. However, \texttt{sk-admin} can be instructed to replace secrets that already exist. 

Once it creates and stores the secrets in Vault, \texttt{sk-admin} exports them to a configurable destination depending on the deployment mechanism used. For example, \texttt{sk-admin} can generate Kubernetes secrets to support deploying the Tapis services to a Kuberenetes cluster; it can also export the secrets to a text file, which can be encrypted and used for deploying Tapis via Docker Compose. 

After \texttt{sk-admin} runs and generates secrets, the next step depends on the type of site being bootstrapped. For primary sites, the Tenants API must be deployed next. The Tenants service reads its configuration as well as the secrets generated by \texttt{sk-admin} from its environment -- for Kubernetes-based deployments, this means injecting Kubernetes secrets into the Tenants service pod as environment variables. In particular and most critically, Tenants retrieves the public key associated with the administrative tenant for the primary site as well as any associate sites and makes these keys available through an unauthenticated endpoint.

Associate sites never run the Tenants API; instead, once \texttt{sk-admin} runs, the associate site system administrator must manually send the public key for its administrative tenant (generated by \texttt{sk-admin}) to the primary site administrator who in turn adds it to its \texttt{sk-admin} configuration. At this point, \texttt{sk-admin} must be re-run on the primary site and the Tenants service restarted in order to add the new associate site public key to the tenants registry. 

Next, SK is deployed, followed by Tokens and Authenticator. Note that all of these services, and in particular SK, depend on both secrets generated by \texttt{sk-admin} as well as the registry of tenants made available by the Tenants API, and hence the specific order of service deployment. At start up, SK reads credentials from its environment that allow it to connect to Vault and manage the secrets associated with the tenants owned by the site. The Tokens service depends on SK for the private key associated with the tenants -- however, a chicken-and-egg problem requires the private key associated with the site's administrative tenant to be injected directly into Tokens's environment at start up. Tokens uses this key to sign a JWT for itself, and then it uses this JWT to query SK for the private keys for all other tenants at the site. Conceptually, this private key amounts to a shared secret between Tokens and SK. 

With Vault, Tenants, SK and Tokens deployed, the Authenticator and any other Tapis services may be deployed. No additional deployment dependencies exist for the remaining services beyond each service depending on Tenants for the tenant configurations, Tokens for generating service tokens, and SK for authorization data. 

As an example, when the Jobs service is launched its service password, database and message broker credentials are injected as environment variables into its multiple containers.  Each container calls Tokens with the service password, which Tokens validates by calling SK.  Tokens then signs and returns the Jobs access and refresh tokens (JWTs) valid in the site's administrative tenant.  Like all services, each Jobs container automatically acquires a new access token before the current one expires (4 hours).  

\section{Challenges} 
In this section we highlight some challenges created by the federated, microservice architecture used by Tapis. Each challenge required new techniques to overcome.

\subsubsection{Preserving the Original Requester in Service Requests}
Most Tapis requests, whether made by a user or a service, require authentication using a JWT passed in the \texttt{X-Tapis-Token} HTTP header. When a service makes a request to another service, it uses a JWT representing itself, but when this service request is being made as part of fulfilling an end-user request, the identity information about the end-user must also be transmitted. Consider the example of user \textit{jdoe} making a request to Jobs to execute an app. User \textit{jdoe} includes her JWT to the original request to Jobs. To fulfill this request, Jobs makes a request to Systems to retrieve the system definition. Jobs includes a JWT representing itself in the request to Systems, but Systems needs to know the identity of the requester, in this case, \textit{jdoe}, for authorization checking and for resolving dynamic system attributes that depend on the specific end-user's identity. 

To allow the identity of the original requester to be transmitted across service requests, Tapis uses two additional headers: \texttt{X-Tapis-User} and \texttt{X-Tapis-Tenant}, referred to as ``On Behalf Of" or OBO headers. As the names suggest, these headers take values equal to the username and tenant of the original requester, respectively.  Only service requests can set these headers -- if a user request sets either of the OBO headers, it will be rejected. 

\subsubsection{Service Request Routing}
Associate sites depend on the primary site for services they do not run, which means that requests can be forwarded between sites.  When a user request is received at a tenant's host, the HTTP proxy extracts the target service from the request's URL path.  As mentioned previously, the proxy then routes the request locally or to the primary site based on the site's service configuration.  Proxy configurations must be kept synchronized with Tapis site configurations which, thankfully, rarely change.  The upshot is that request routing can cause JWTs created at one site to be validated at another.   

\subsubsection{Service Request Validation}
To ensure cross-site security, receiving services validate JWT content by checking that the following conditions are met:
\begin{enumerate}
    \item The JWT was signed using the private signing key of the tenant specified in the JWT's \textit{tenant\_id} claim.
    \item The receiving site is configured to run the targeted service.
    \item If the request targets SK or Tokens, then the JWT tenant must be owned by the receiving site.
    \item If the receiving site is the primary site and the JWT tenant's owning site is an associate site, then the associate site CANNOT be configured to run the targeted service. 
    \item If the receiving site is an associate site, then the sending site must be the primary site.
    \item If a \textbf{user JWT} is received, then
    \begin{enumerate}
        \item neither the \textit{on-behalf-of} user nor tenant header fields can be set, and
        \item the JWT tenant cannot be any site's administrative tenant.
    \end{enumerate}
    \item If a \textbf{service JWT} is received, then
    \begin{enumerate}
        \item the \textit{on-behalf-of} user and tenant header fields must be set, and
        \item the JWT \textit{target\_site} claim must be set to the receiving site, and
        \item the JWT tenant must be the admin tenant the of site (1) that owns the \textit{on-behalf-of} tenant and (2) where the sending service is configured to run.
    \end{enumerate}
\end{enumerate}
Only service JWTs contain the \textit{target\_site} claim, which means that the sender of a service-to-service request must know the receiving tenant's site.  This further implies that services running at the primary site need a unique JWT for every site, and that services running at associate sites need a JWT for their site and the primary site.  We've seen how user requests are forwarded from associate to primary site, also note that service requests can be forwarded in either direction depending on where communicating services are located in a tenant.  Shared library code used by all services keeps JWTs cached and refreshed, and selects the appropriate JWT for outgoing requests based on target site.

%% file: 04_performance.tex
\section{Performance Assessment}  
\label{section:perf}
In this section, we evaluate SK performance, key to the Tapis security architecture described in section \ref{sec:tapis-sk}.
The evaluation consists of load testing certain SK endpoints in different scenarios.  A scenario is defined with respect to the number of permission objects in SK, the simulated number of concurrent users sending requests, the SK endpoint, and the wait time between requests. The goal is to gauge the scalability of the SK Permissions API, specifically, the \textit{isPermitted} end-point. 

\subsection{Experimental Setup}
\subsubsection{Data Generation}
 To mimic a real-world scenario, we defined five user roles: scientists, developers, project managers, collaborators, and the public. Each user is granted permissions to file paths in a system based on their roles. 
 

For each Tapis test system, we created 1000 path permissions in SK. To test scalability, our scenarios comprised [1K, 10K, 25K, 50K, 100K] permissions, which in the largest case implies 100 systems each with 1000 permissions.  
\subsection{Load test design}
Our client machine had 32 CPUs, an Intel\textsuperscript{\textregistered} Xeon\textsuperscript{\textregistered} E5-2660 CPU @ 2.20GHz and 128 GB RAM. We designed our test using Locust v2.14.2, an open-source load tester \cite{locust}. On the client machine, Locust generated HTTPS requests to SK's \textit{isPermitted} endpoint with parameters \{\textit{tenant}, \textit{username}, \textit{permission string}\}. We set the wait time between requests to be randomly assigned from the (0.01, 0.1) second range.

\subsection{Evaluation Results}
\begin{figure}%
    \centering
    {{\includegraphics[width=8cm]
    {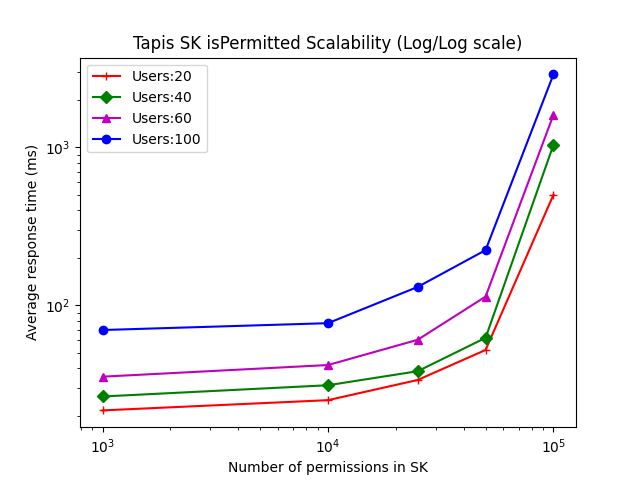}}}%
    \caption{Tapis SK API isPermitted load testing result }%
    \label{fig:is_permitted}%
\end{figure}

Table \ref{tab:exp-scenario} shows the results for scenarios with different total permissions and for either 20 or 100 concurrent users. Figure \ref{fig:is_permitted} plots the results, including those for 40 and 60 users not shown in the table.  Request pacing led to zero observed failures.

The key takeaways from our benchmarks are:
\begin{enumerate}
\item As the number of permissions in SK increases, the average response time for \textit{isPermitted} requests increases sublinearly up to 50K, which is good. Between 50K and 100K, the average response time increases by an order of magnitude, which indicates a bottleneck.

\item As the number of permissions in SK increases, the requests per second decrease.  Again, the effect is most dramatic when permissions increase from 50K to 100K.

\item As the number of concurrent users increases, the average response time increases. This is expected behavior since SK can only process requests at a certain rate.  Interestingly, increasing wait times to 1-2 seconds reduces response times (see below).

\item Minimum and maximum response times represent the range of response times for each scenario. The minimum, average, and maximum response times increased significantly in the 100K case.

\item 99.9 percentile response times represent the values within which 99.9\%  of all requests get processed. For example, in scenario 4, 99.9\% of requests are processed within 120 ms, reasonable for web calls. 
\end{enumerate}

When we increased the wait time to 1-2 seconds (not shown), we saw a moderate response time of 500ms for 70 concurrent users with 100K permissions.  With 50K permissions and 0.01-0.1 second wait times, 100 concurrent users experienced a 224ms average response time.

Investigating the spike in response times in 100K scenarios, we identified database queries that returned approximately 17K permissions, each of which incurs parsing and regular expression matching, to be the  major factor. The details of these experiments are omitted due to space limitations, but the key point is that we rarely see users granted even 50 permissions, \textit{so the load tests stressed the system several orders of magnitude beyond our current needs}.

We also compared response times for Systems \textit{getPermissions} calls from a primary site and from an associate site located across an ocean. Systems only runs at the primary site, so associate site requests incurred a round trip 265ms penalty for the extra hop in our routing algorithm as expected.      

\begin{table*}
\centering
\caption{Evaluation Results}
\label{tab:exp-scenario}
\begin{tabular}{|p{1.2cm}|p{1.5cm}|p{1.0cm}|p{1.5cm}|p{0.9cm}|p{1.3cm}|p{1.3cm}|p{1.3cm}|p{1.3cm}|p{1.3cm}|p{1.3cm}} \hline
\textbf{Scenario} & \textbf{\#Permissions} &\textbf{\#Users} &\textbf{\#Requests} &\textbf{RPS} & \textbf{Min Resp (ms)}&\textbf{Avg Resp (ms)} & \textbf{Max Resp (ms)}&  \textbf{75\%ile Resp (ms)} &\textbf{99.9\%ile Resp (ms)}  \\ \hline
1 & 1K &20 & 28477 & 246 & 12 & 21.64 & 242 & 24 & 68 \\ \hline 
2 & 10K &20 & 31113 & 241 & 16 &25.13 & 227 & 27 & 71  \\ \hline 
3 & 25K &20 & 	23761 & 218 & 24 & 33.73& 209 & 36 & 100\\ \hline 
4 & 50K &20 & 16525 & 180 & 36 &52.23 & 214 & 57 & 120  \\ \hline 
5 & 100K &20 & 15256 & 35 & 284 & 501.49 & 1452 & 540 & 1200 \\ \hline
6 & 1K &100 & 161962 & 754 & 13 & 69.85 & 3163 & 92 & 320  \\ \hline 
7 & 10K &100 & 170035 & 705 & 16 & 77.17 & 3194 & 97 & 320  \\ \hline 
8 & 25K &100 & 100119 & 500 & 24 &130.90 & 3112 & 170 & 490  \\ \hline 
9 & 50K &100 & 52758 & 329 & 37 & 223.80 & 1491 & 280 & 990  \\ \hline 
10 & 100K &100 & 20370 & 33 & 294 & 2896.89 & 18133 & 3600 & 14000 \\ \hline
\end{tabular}
\end{table*}

%% file: 05_usecase.tex
\section{Use Cases}
\subsection{Hawaii Climate Data Portal}
Climate scientists require large datasets to build predictive models to study the impacts of climate change and understand issues such as optimal land use and water recharge rates. Climate stations installed across the Hawaii islands collect measurements of variables such as temperature, humidity, rainfall, etc. To make use of these data, a team of researchers at the University of Hawaii Manoa (UH) are building an automated data collection and analysis platform using Tapis. UH runs a Tapis associate site on premise with the Streams API, among others. Climate stations stream raw measurement data in real-time to the UH Tapis installation. Quality assurance checks, such as anomaly detection, execute as the data arrive to detect issues with the collected data. Analysts package the quality assurance scripts into Docker containers registered as Tapis functions which the platform executes automatically in response to receiving new data. Additional programs, registered as both Tapis functions and applications, run at certain intervals to produce detailed climate maps.

While much of the analysis executes on local resources at UH, some computations require advanced hardware or large storage that is either in high demand or unavailable entirely on site. For example,  state-of-the-art isotope-enabled cloud resolving models, using Lagrangian components that tracks the positions of billions of particles at every simulation step,  improve understanding of processes influencing water isotopic composition of water vapor in the atmosphere and require high performance large scale storage and hundreds of CPUS per model run. 
The Hawaii Climate Data Portal (HCDP) provides comprehensive access to the raw data and downstream data products in a central web resource. Using HCDP, researchers across the United States and internationally access, interact with and visualize data directly in the browser. 

\subsection{NASA NEID and Tapis Data Pipeline}
As part of the NEID project, NASA JPL and partners deployed a large spectrometer on the top of Kit Peak National Observatory in Arizona. The NEID spectrometer makes precise measurements of the wobbles of nearby stars to discover exoplanets, that is, planets orbiting stars other than our sun \cite{neidOveriew}. The NEID spectrometer generates a level 0 data file, called a FITS file, for each exposure captured, including scientific exposures and calibrations. Each file contains the raw 2D spectrum for each the 16 readout channels.

The NEID project developed a series of 10 data analysis programs in Python, IDL and C to produce level 1, level 2 and additional data products from the raw exposure data, collectively referred to as the NEID analysis pipeline. These programs require on the order of 1024 GBs of RAM per run. The pipeline must run to completion within 24 hours of data collection to be useful. To accomplish this, the team built an automated data analysis framework on top of Tapis consisting of the following high-level steps:
\begin{enumerate}
    \item Raw exposure data (i.e., level 0 files) are transferred from the NEID spectrometer on Kit Peak to a storage resource residing at Caltech. This is the primary (source) copy of the data.
    \item Exposure data are then copied from Caltech to the Texas Advanced Computing Center (TACC) at the University of Texas, Austin for processing.
    \item A multi-step analysis processes sets of level 0 files, which execute on various machines at TACC.  
    \item A Tapis transfer moves the resulting data products (level 1, level 2, etc.) from TACC back to Caltech.
\end{enumerate}
In addition to automating the steps above, Tapis captures metadata about pipeline execution, including size and time of data transfers, job execution times, resource utilization on different computing systems, etc. Tapis also schedules jobs to run on alternative systems at TACC when primary systems are unavailable, such as during maintenance windows. 

%% file: 06_related_work.tex
\section{Related Work}

Apache Airavata \cite{airavata} is a framework for scientific computing that uses a security manager to intercept all requests and communicate with an OAuth2 server to validate access tokens, check roles and manage identities.  By contrast, Tapis services validate access tokens without making a network call.  Tapis uses its own authorization service (SK) and does not have its own identity provider.  Open OnDemand (OOD) \cite{ondemand} specializes in providing graphical interfaces to users at HPC centers through the clever use of proxies and POSIX access controls.  Tapis also relies on host access controls, but layers its own roles, permissions and sharing model on top of them.

The Globus Auth service \cite{Globus-Auth} provides APIs for research data management and enables researchers to access data across institutions and via multiple, linked identities. Globus is proprietary software and source code is not publicly available. Certain features require a paid subscription. SciTokens and the related SciAuth project provide standards for credential format based on the JWT specification \cite{SciTokens}\cite{SciAuth}. Adoption of these standards could improve interoperability of credentials across service providers, something being considered in Tapis. 

%% file: 07_conclusion_futurework.tex
\section{Future Work and Conclusion}
Future work includes the development of SK \textit{projects}, which are shared workspaces that associate a set of users with a set of resources.  We also plan to improve support for federated identities, multi-factor authentication (MFA) and token-based login on HPC systems.  Federation includes more support for Globus Auth and InCommon.  Login tokens that contain MFA information would strengthen security when Tapis logs in with no human in the loop.  Finally, we'd like to formally prove the security characteristics of our protocols and implementation. 

We have shown how Tapis addresses security concerns in multi-site, multi-institution deployments while preserving the flexibility of a microservice architecture.  Our tenancy model, with its site ownership and administrative tenants, separates user groups and shares capacity across sites.  Tapis authentication can be used as-is or customized to fit into almost any environment.  Secrets can be restricted to a site and our unique token routing algorithm allows trust between services at different sites.  We have shown how roles, permissions and distributed shared contexts allow for efficient authorization and micro-benchmarking shows that security processing is not a bottleneck.  Finally, two science use cases demonstrate how Tapis is being applied to real world problems.    